# Shapley Value Is not Applicable To Network Access Pricing.

MICHAEL NWOGUGU
Address: P. O. Box 996, Newark, NJ 07101, USA.
Email: mcn2225@aol.com; mcn111@juno.com.

**Abstract**
Although Game Theory principles have been used extensively in developing analytical models for access pricing in Internet networks, transmission pricing in electric power networks and access pricing in telephone/cable networks, most of the existing literature omits the effects of illegal filesharing. This article shows that: 1) Shapley Value analysis is generally inaccurate for most situations, and 2) Shapley Value analysis are generally not applicable to network analysis.

## Introduction

File sharing and illegal downloads of content have resulted in millions of dollars of losses for many companies and substantial lawsuits by trade groups and entertainment companies. Illegal downloads affect the economics, profitability and business models of companies in many industries such as entertainment, education, travel, investments/finance, and any business where knowledge has value. In the US, the Napster case illustrates some of the policy, technological and economic issues inherent in systems for downloading content. Golle, Leyton-Brown, Mironov & Lillibridge (________); Clark & Tsiaparas (2002). See: www.news-service.stanford.edu/news/2004/march17/fileshare-317.html.

## I. Existing Lierarture

Some authors/researchers have attempted to apply the Shapley Value (and related theories) to access pricing, and to allocation of revenues/costs in networks. Ma, Chiu, Lui, Mira & Rubenstein (2009). Cao, Shen, Milito & Wirth (2002); Nigro & Abbate (2009); Kattuman, Green & Bialek (2004); Jia & Yokoyama (2003); Bakos & Nault (1997); Cao, Shen, Milito & Wirth (2002); Morais & Lima (2007). Angel E, Bampis E, Blin L & Gourves L (2006). Altman & Wynter (2004).

Several Authors/researchers have explained the inaccuracies and problems inherent in the Shapley Value and related theories. Tick, Yap & Maher (________) stated that the Shapley Value algorithms don't relate to the constraint space, but rather, are based only on the coalition payoffs; and thus Shapley Value type analysis can produce an allocation even for an inconsistent problem without any rational agreement point. Fatima, Wooldridge & Jennings (______) stated that: **a)** the problem of determining the Shapley Value is #P-complete, and **b)** the Shapley Value provides an allocation only with a limited degree of certainty (there is an "uncertainty" problem). Roth (1977) showed that a game's Shapley Value is equal to its utility if and only if the players' preferences are neutral to (not affected by) both "ordinary risk" and "strategic risk"; and otherwise, the Shapley Value is not an accurate allocation tool or decision-making tool. Sastre & Trannoy (2001) found that the Shapley Value rules don't satisfy the principle of independence of the aggregation level (the marginal contribution of a player depends on how other components of the allocated resource are treated). Kirman, Markose, Giansante & Pin (2007) found various problems inherent in the Shapley Value function. Lima, Contreras & Padilha-Feltrin (2008) found that were no feasible cores in cooperative games for allocation of costs in networks. Roth & Verrecchia (1979) explained the limitations and inconsistencies inherent in the original Shapley Value Theory; and stated that Shapley Value analysis does not account for "Fairness", and then they introduced a new explanation of Shapley Value theory. Bakos & Nault (1997) explained the conditions under which ISPs and certain coalitions will invest in network infrastructure – their analysis is critical to understanding the need for and potential success of coalitions among ISPs. Biczok, Kardos & Trinh (2008) show that Shapley Value analysis does not account for brand loyalty and QoS.

Some Authors have attempted to use Stacklberg Bargaining analysis for Network pricing. See: Korillis, Lazar & Orda (1997); Shakkotai & Srikant (2006). Most of the critiques and weaknesses of Shapley Value analysis, also apply to Stackelberg Game analysis; and thus the inaccuracies and problems summarized above in Tick, Yap & Maher (________); Fatima, Wooldridge & Jennings (______); Roth (1977); Sastre & Trannoy (2001); Kirman, Markose, Giansante & Pin (2007); Lima, Contreras & Padilha-Feltrin (2008); Roth & Verrecchia (1979); Bakos & Nault (1997); and Biczok, Kardos & Trinh (2008), also apply to Stackelberg Bargaining analysis. He, Prasad, Sethi &Gutierrez (2007).

## II. Mechanism Design Theory Is Inaccurate.

Shapley Value and Stackelberg Bargaining as applied to Interconnection pricing are essentially Mechanism Design Theory. Mechanism Design theory is generally inaccurate. Mechanism Design Theory ("MDT") is inaccurate and impractical. MDT is the body of theory pertaining to economic mechanisms that are often defined by rules, norms, time, medium and location.[1] The literature on MDT [2] has some major gaps and inaccuracies, some of which are explained as follows. Myerson R (2002). Myerson (1983); Friedman & Oren (1994); Larson & Sandholm (July 2005); Sabzposh (_______).

MDT Errornously assumes that all agents truthfully disclose their preferences, and that all agents disclose their preferences at the same rate and at the same time. MDT Does not account for the value that the agent or principal or participant gains by withholding information about their preferences.

MDT errornously assumes that each mechanism is fair and un-biased; but in reality, even completely automated mechanisms have biases (that arise from programming errors, computer learning patterns, and or customer usage patterns). Most mechanisms involve some human intervention and or human processes, and existing MDT does not account for human biases and processes such as Altruism, Regret, aspirations, etc., both in the participants and in the humans that are involved as part of the mechanism.

---

[1] See: http://nobelprize.org/nobel_prizes/economics/laureates/2007/ecoadv07.pdf.
[2] *See*: Myerson R (2002). Optimal Coordination Mechanisms In Generalized Principal Agent Problems. *Journal Of Mathematical Economics*, 10(1):67-81.
*See:* Myerson R (1983). Mechanism Design By An Informed Principal. *Econometrica*, 51(6):1767-1797.
*See:* Friedman E & Oren S (1994). The Complexity Of Resource Allocation And Price Mechanisms Under Bounded Rationality. *Economic Theory*, 6(2): 1432-1479.
*See:* Larson K & Sandholm T (July 2005). *Mechanism Design And Deliberative Agents*. Presented at AAMAS 2005, July 2005, Utrecht, Netherlands (ACM).
*See:* Sabzposh A (_______). *Routing Without Regret: On Convergence To Nash Equilibria Of Regret Minimizing Algorithms in Routing Games*.

Existing MDT does not account for varying levels of "confidentiality" of agents' information – rather, MDT erroneously assumes that a binary situation exists in which information is either publicly known or is completely private. Existing MDT does not incorporate the effects of government regulation on agents and on the mechanism; and does not account for the un-constitutionality/constitutionality of mechanisms. MDT erroneously assumes that all agents are "rational" and self-interested; but there can be many reasons for agents' irrationality and or agents' propensity to act for the benefit of the society or for the benefit of an interest group. MDT erroenously assumes that there is always some minimum level of uniformity of agents' preferences; but on the contrary, Agents' preferences vary widely.

MDT does not account for differences in agents' information-processing capabilities. MDT erroenously assumes that each mechanism is monolithic and static in time, space and expense – in reality and on the contrary, some mechanisms are dispersed in space (various locations) and time (requires participation, revisions of knowledge and various disclosures at various times) and expenses (the costs of participation in the mechanism varies).

MDT erroneously assumes that in mechanisms, monitoring costs, compliance costs, switching costs, access costs, decision costs (the costs of contemplating a decision) and sanctions (for non-compliance with the mechanism) are minimal or non-existent. In reality, these types of costs are monetary/physical and or non-monetary/psychological and can have significant effects on the efficiency of mechanisms.

MDT erroneously assumes that agents have quasi-linear utility functions and are risk-neutral; but in reality, agents' attitudes towards risk vary dramatically and depend on many factors. Furthermore, agents' utility functions are more likely to be non-linear (and not linear) because the agent will react to the various dimensions of the mechanism (economically, psychologically and socially), and will also react to the prospect of there being other participants, and also react to perceived opportunity costs, in addition to his/her normal utility function.

MDT erroneously assumes that the social choice functions inherent in mechanism designs have linear "Benefit Effects"; where a Benefit Effect is defined as the economic gain or loss of social welfare across all agents and across the society/economy, as the mechanism functions during a specified time interval. Hence, the Benefit Effect is defined with respect to time and to the entire economy. Contrary to MDT, Benefit Effects are likely to be non-linear because agents characteristics vary in terms of wealth, utility functions, risk aversion, time horizon, preferences, etc.; and the economy is not static, and changes in various elements of the economy are not discrete; and finally, not all eligible agents or permitted agents or financially-capable agents will participate in the mechanism.

MDT errorneously assumes that the social choice functions inherent in mechanism designs have uniform and same "Impact Effects" across all agents; where an Impact Effect is the magnitude of the monetary and non-monetary impact of the mechanism on all agents. MDT erroenously assumes that all social choice functions inherent in mechanisms have linear effects on agents' utilities and participation strategies

MDT erroneously assumes that all eligible, financially capable and permitted agents will participate in the mechanism, and will participate at the same time. MDT erroenously defines the success/efficiency of mechanisms primarily in terms of utility; but this approach does not sufficiently incorporate other elements and results of mechanisms – such as psychological gains/losses, emotions, social capital, reputation effects, etc.. Futhermore, as used in Mechanism Design Theory, utility is relatively static. McCauley (2002)[3] states that there are several problems in the use of utility. Most MDT are based on equilibrium as a relevant 'state' and as an objective; and the concept of equilibrium is "static". In reality true equilibrium does not exist, and cannot be achieved in mechanisms due to: a) continous changes in agents' preferences, information processing capabilities, time constraints, wealth, access to information, etc., b) transaction costs and opportunity costs, c) mental states of agents, d) government regulations and or industry standards/practices, e) agents' varying reactions to incentives over time.

MDT errorneously assumes that each agent's and all agents' preferences are static over time; and mechanisms are preference formation-independent (ie. the mechanism does not affect the agents' processes of forming their preferences). In reality, most mechanisms are interactive, and the agent's preferences change over time as he/she interacts with both the mechanism and other agent-participants and non-participants. In most MDTs, Mechanisms are defined and designed only in terms of agents' preferences, public actions, and private actions. This approach does not incorporate the effects of agents' reactions to incentives, and values of hidden information to agents, and agents' information processing capabilities, the mechanism's information processing capabilities, regulation and government enforcement. Contrary to MDT, the set of all possible preferences of agents is not finite. Within this context of mechanisms and group action, the definition of 'finite" should be based on achievability, and not on mathematical ranges.

---

[3] *See*: McCauley J (2002). Adam Smith's Invisible Hand Is Unstable: Physics And Dynamics Reasoning Applied To Economic Theorizing. *Physica A Statistical Mechanics And Its Applications*, 314(1-4):722-727.

MDT errornously assumes that the mechanism is removed from, and distinct from the agents – in reality, the agents typically form a major part of the mechanism (as in auctions, online file sharing networks, multiple listing systems, etc.). MDT errorneously assumes that the mechanism's main role is either allocation and or coordination. In reality many mechanisms serve other economic and non-economic purposes (some of which are un-intended) such as: a) psychological reassurance (voting, auctions, etc.), b) information dissemination, c) comparison – which increases social welfare by reducing overall agents' search costs, d) entertainment. MDT erroneously assumes that mechanisms can be deliberation-proof (in equilibrium, all agents don't have any incentive to strategically deliberate). In most existing mechanisms, agents deliberate while using the mechanism.

### **III. Interconnection Fees Can Be Irrelevant In Network Design.**

The existing literature on network access pricing un-necessarily focuses on Interconnection Fees (Peering Fees and Transiting Fees) as the main element for the improvement of networks. The main method of analysis has been Game Theory which has been shown to have major weaknesses. Under established principles of Game Theory and Dynamical Systems analysis, the Interconnection Fee is almost irrelevant if any of the following conditions exist:

1) The Interconnection Fee is a very small component of the total average per-packet (per unit) end-to-end cost of transmitting packets in the ISP's network or in the combined coalition members' network.

2) For purposes of financial reporting, Network transmission costs (amortization, deprecation of network infrastructure, Interconnection fees and bandwidth) must be expensed in the period incurred.

3) The market is protected or regulated – eg. by government regulations. The government may also decide to fix or reduce or limit Inter-connection fees in order to promote competition.

http://wirelessfederation.com/news/5212-romanian-mobile-operators-challenge-interconnection-fees/.

http://www.telecomsinsight.com/file/62892/interconnection-fees-lowered-by-swisscom.html. ITU (2007).

http://www.wgig.org/docs/book/EB_JF.html. http://www.nytimes.com/2009/08/17/technology/17iht-terminate.html.

4) In most jurisdictions, Antitrust laws prohibit the direct or indirect formation of coalitions (especially coalitions of the type envisaged in Shapley Value or Stackleberg Game analysis).

5) ISPs can terminate Peering/Interconnection Agreements at-will or with minimal termination costs.

6) For any distance *d* in any given region *r*, ISPs can quickly and cheaply re-configure their physical networks (routers, switches, etc.). This condition is feasible in many wireless networks, and particularly with the new generation of wireless networks where each mobile device is a node in the network and also carries/transmits third-party traffic signals.

7) The cost of building wireless networks is lower than the cost of land-based networks.

8) There is no required minimum "contribution" to a coalition by each ISP-member (where "contribution" is the required minimum size of the ISP's network). In such cases, an ISP with minimal or no physical network will be able to obtain a share of the coalition's revenues from the transmission of its packets.

9) Quality of Service (QoS) is a primary/dominant factor in ISP's decisions about Interconnection Agreements.

10) In most markets, of the two types of Interconnection Agreements ("Peering" and "Transiting" Agreements), Peering Agreements account for the super-majority (eg. more than Seventy Percent) of all Interconnection Agreements, and Peering Agreements do not result in any additional costs for the ISP (each ISP agrees to carry the other ISP's traffic for free).

Note that Interconnection fees are only one component of the ISP's costs and other cost components include: a) the non-transmission costs – such as administration, marketing, compliance, etc..; b) physical maintainance costs; c) costs of congestion. Thus, Interconnection fees may be a non-material component of the ISP's cost structure; and in such circumstances, the ISP's propensity to join any coalition will not be driven by interconnection fees, but may be influenced by other factors such as reputation, brand equity, Anti-trust laws, other ISPs' "Contributions", the subject ISP's planned growth, innovation and availability ofc cash.

The ISP incurs various types of costs in order to transmit content to customers, and these costs are: a) administrative costs, b) marketing costs, c) infrastructure maintainance costs, d) capital expenses – network infrastructure costs, e) variable transmission costs- bandwidth, professional staff, Interconnection fees, f) compliance costs, etc..

## IV. The "Shapley Value" Is Inaccurate And Cannot Be Used In The Analysis of Access Pricing Or For The Allocation of Revenues/Costs In Networks.

Within the context of Interconnection Agreements, the Shapley Value can be accurate and relevant if and only if certain conditions exist - the following are the conditions required for Shapley Value to be valid, and why these conditions are false and don't exist.

**Theorem #1**: *The physical structure of the Internet is never constant*.

*Proof*: In reality, although the physical connections of ISP's networks may remain fixed in time and space, the true structure of the network (the actual transmission pathways, and hence the "Network") is not constant, and changes instantenously, depending on the locations of the sender-receiver pairs. In almost all instances, each packet has more than one feasible path. Furthermore, in land-based networks the cable used in transporting packets typically has more than one section of wires so that for each distance *d* in any cable, there is more than one feasible transmission pathway. Similarly, in any wireless network, for any two points in distance *d*, there are many possible transmission paths. ∎

**Theorem #2**: *For any transmission, even if the Originating ISP knows the cost structure and payoff of every other ISP that is required to transport any given packet, optimal allocation of costs or revenues among cooperating ISPs is never feasible.*

*Proof*: The Shapley Value analysis erroneously assumes that each ISP knows the cost structure and payoff patterns of every other ISP that is a member of the coalition. In reality, statutes prevent public disclosure of cost and pricing information of ISPs; and statutes also prevent ISP from sharing information among themselves. Active competition among ISPs will almost always preclude their voluntary disclosure of their cost/price data to each other. ISPs can guess about their competitor's cost/price information and can recruit their competitors' staff in order to give them such information. However, network transmission costs are constantly changing not only because of the variations in the number and quality of feasible transmission paths, and differences between Full-costing and Variable-costing, and the dynamic nature of wireless portions of networks, but also because of the following factors: i) entrance and exits of ISPs into markets, ii) inflation, iii) calculation of depreciation, iv) changes in capital expenditures, v) changes in marketing costs.

Assume a "second best-case" scenario where the competitors' cost/price data are fully observable to the Originating ISP (O-ISP), and there is a coalition, but coalition-members dont coordinate/unify their routing systems/protocols. The O-ISP can calculate a rough estimate of costs from point $P_1$ to point $P_2$, but O-ISP cannot dictate to other network members about when and how to route the packet and to which "third"-ISP to transmit the packet – and thus, each network member will typically be selfish despite coalition agreements. This state is referred to as *Coordination-Control Neutrality* and it prevents any optimal allocations. Assume a "first best-case" scenario, where all ISPs in the network cooperate to try and determine the most efficient routing path, and also disclose their cost/price data to each other. As mentioned, such cooperation will be deemed illegal in most jurisdictions. Even if such cooperation is feasible, there can never be an optimal allocation because of the time frame of the required decisions. Assume that the transmission goes though several ISPs (O, A, B, C, D.......N) sequentially, for distances $d_1, d_2, d_3, d_4, ..........d_n$ in time intervals $t_1, t_2, t_3, t_4, .........t_n$. Note that A represents the set of all feasible ISPs that can handle the transmission over $di_2$ in $tl_2$; and B represents the set of all feasible ISPs that can handle the transmission over $d_2$ during $t_2$, and so on for all subsequent transmissions. Hence, the time intervals $t_1, t_2.....t_n$ are stochastic and un-known in prior periods and the choices A, B, C......N are also unknown because of the changing nature of cost/price pairs of information. For efficient cooperative allocations, each transmitting ISP (O, A, B, C, ........N) will have to provide instantenuous cost/price pairs of data ($O_{cp}$, $A_{cp}$, $B_{cp}$, $C_{cp}$,.......$N_{cp}$), for each distance $d_1, d_2, d_3,.........d_n$. Hence for A to transmit to B during time $t_2$, A must have at least $A_{cp}$ and $B_{cp}$, and $C_{cp}$ at the beginning of $t_2$; and for B to transmit to C during time $t_3$, B must have at least $B_{cp}$ and $C_{cp}$, and $D_{cp}$ at the begining of $t_3$; and so on for subsequent transmissions. By the time the packet moves from B to C, $D_{cp}$ and $E_{cp}$ and $F_{cp}$ have changed, and a new allocation must be made. The sensitivity of the optimality of allocations to changes in the cost/price pairs is very high because of the nature and volumes of the transmission of packets (the components of each ISPs total transmission costs and prices can vary dramatically across different ISPs, and are explained above). Because these total transmission costs (TTC) and prices are difficult to calculate and the micro-time (per-millisecond) cost data required to calculate them are available infrequently (at best, on a daily or hourly basis), there is sufficient accurate data for determining any optimal allocation of costs or revenues among a cooperative network of ISPs. Secondly, because most ISP inter-connections are free, introducing the voluntary or mandatory exchange of cost/price data among ISPs is likely to create substantial dis-agreements over fairness of allocations, and or will also affect (reduce) each ISP's willingness to invest capital to build physical or wireless networks. This is because each ISP calculates its return-on-capital and marketing ROI differently. Such Disclosure-Influenced Free Riding will in turn, limit the growth of the coalition's overall network and preclude or eliminate any efficiency of the allocations of costs/revenues among the ISPs. ∎

**Theorem #3**: *For every packet and for each time interval t, and for any distance d in the coalition's combined network, each ISP (originating ISP or transmitting ISP) cannot choose an optimal path*.

    *Proof:* An ISP cannot control the routing choices of other ISPs. In reality the selected path for each packet is determined by pre-established automated algorithms of each ISP, and these algorithms don't always adapt to changes in the network physical structure or transmission-path(s) – which was referred to above as

*Coordination-Control Neutrality.* Furthermore, because each ISP is not able to obtain very accurate information about historical or future packet loss, its impossible for the ISP to plot a "optimal transmission path". Its highly unlikely that any ISP will permit other ISPs to control or overly influence its routing algorithms even in a cooperative network. Antitrust laws will almost surely prevent such coordination. Routing algorithms are complex and are sensitive to changes in network characteristics and the level of human input. The only feasible way that ISPs in a cooperative network can safely influence each others routing algorithm is if they all use the same commercially-available Routing Algorithm - which is highly unlikely. The second-best alternative is to provide a system of cash incentives to convince each ISP to allow other ISPs to influence their Routing Algorithms. The properties of any such incentive (I) are as follows:

**a)** *Consistency and Time Invariance*: the ISP cannot deviate from the incentive mechanism Q.

**b)** *Scope limitation* – Q is limited only to participating ISPs. This presents a huge problem because coalition members cannot control the growth of non-member ISPs. In some markets, Coalition members must connect with non-members of the coalition, in order to complete transmissions.

**c)** *Equivalency* - $\int_0^t Q_i \, dt > \int_0^t \Sigma \, O_i \, dt$, where $O_i$ is the value of the ISP's opportunity costs of complying with the incentive mechanism in every transmission. $Q_i$ is the value of the incentive mechanism for the ISP for transmission i.

d) *Additivity*: for any transmission over $d_i$.

Because any given ISP cannot "choose" an optimal path in the Network, Shapley Value analysis is not applicable. ∎

**Theorem #4**: *For Any Transmission Over Any Distance (d), The ISP's Share of Revenues (or Share of Transmission Costs) Generated By The Coalition Is Not Always Non-Negative (Can Be Negative).*

*Proof*: The Shapley Value function errorneously assumes that the ISP's share of revenues generated by the coalition is always non-negative, and that the ISP's share of costs generated by the coalition transmission is always non-negative. On the contrary, there are some instances where the ISP's share of total transmission revenues generated by the coalition is negative (less than zero) or should be negative, and these include the following:

a) When the ISP's transmission results in excessive packet loss such that the entire information-unit has to be re-transmitted.

b) Where the ISP earns non-coalition third-party benefits for carrying the packet, but is effectively subsidized indirectly by coalition members.

c) Where given specific transmission algorithms (and congestion control algorithms) that are implemented by coalition member ISPs, ISP-A's transmission of a packet increases the probability that certain other types of packets will be sent by both coalition member and non-members to, and transmitted through ISP-A's network.

d) Where transmitting a certain packet through the ISP's network exponentially increases each subsequent coalition member's costs of carrying the same packet to its destination.

e) Where transmission by an ISP results in addition of data to the packet which then increases the required minimum bandwidth and costs for further transmission by all other coalition members. ∎

**Theorem #5:** *For any time interval t, and for any distance d in the Coalition's networks, the marginal contribution of each coalition-member ISP is not Always Directly Proportional to: 1) Distance, or 2) Bandwidth Consumed.*

*Proof*: The Shapley Function erroenously implies that for any time interval t, and for any distance d in the Coalition's networks, the marginal contribution of each coalition-member ISP is directly proportional to: 1) distance, and 2) bandwidth consumed. This is not always true. For example, with regard to distance, the marginal contribution of a coalition-member ISP that transports a packet for 100 meters in an extremely high density city-block with significant internet congestion, is greater than the marginal contribution of an ISP that transports a similar packet for 1,000 meters in a low-density rural area with no internet congestion. ∎

**Theorem #6**: *For any time interval t and for any distance d in the Coalition's networks, the number of coalition members is not always non-negative.*

*Proof*: The Shapley Function erroenously implies that for any time interval t and for any distance d in the Coalition's networks, the number of coalition members is not always non-negative. An ISP can be deemed a "negative" member of a coalition when any of the following conditions exist:

a) X% of the ISP's network is physically proximate to the networks of coalition members even though the ISP is not a coalition member, and such proximity can reduce the coalition's transportation burden by Q%.

b) The ISP has no physical networks or wireless networks but enters into interconnection agreements with coalition members (or the ISP has minimal physical networks – about 5% of the average network size. ∎

**Theorem #7**: *For any time interval t, and for any distance d in the Coalition members' networks, the marginal Rate of transmission of packets ($R_t$) (the distance that the packet travels for each additional unit of bandwidth consumed) Does Not remain Constant.*

*Proof*: The marginal rate of Transmission of packets varies constantly in most networks for various reasons including traffic/congestion, size/diameter of the network cable, physical obstructions in wireless networks, type of protocol, type of . With such varying $R_t$, each coalition member's contribution cannot be measured by any standard unit, and hence given the Shapley Value theory and formula, any allocation will not be fair or stable. Thus, the Shapley Value Function is inapplicable to access pricing. ∎

**Theorem #8**: *For any time interval t, and for any distance d in the Coalition members' networks, the marginal utility of substituting an ISP in the transmission path (the increase in utility gained from substituting an ISP in the Transmission Path) Does Not Remain Constant.*

*Proof*: Clearly, both the originating-ISP's and the consumer's marginal utility of substituting any ISP in the Transmission Path (path between sender and receiver) changes instantenously due to: a) changes in traffic and volume; b) the size of the pipe; c) the absolute efficiency and relative efficiency of each ISP's network; d) timing; e) transmission fees, and terms of payment of such fees (nature of the contracts); etc.. As explained before, the physical network may be constant but differs from Transmission Paths within it. Thus, Shapley Value Function is not applicable to access pricing. ∎

**Theorem #9:** *The Shapley Value Is Inapplicable Because For each packet, and for any time interval t, and for any distance d in the Coalition members' networks, the end-to-end total Interconnection fee is a Small or Insignificant component of the per-packet total transmission cost; And The Shapley Value Allocations Of Cost And Or Revenues To Coalition Members Will be Un-Feasible When Costs And Or Revenues Exceed Or Fall Below Certain Thresholds.*

*Proof*: In most instances, the Interconnection Fees are not a main component of the total transmission cost (TTC) incurred by the ISP. For any time interval t and for any distance d in the coalition's network, if the ISP's average end-to-end total transmission cost ($R_a$) or Marginal end-to-end Total Transmission Cost ($R_m$) is below a certain dollar threshold (Y), the ISP may not care much about Interconnection Fees and a Shapley Value allocation of cost, and will not have adequate incentives to remain in any coalition of ISPs. Also, if $R_a$ and $R_m$ are above a certain dollar threshold (X), the ISP will face an increasing risk of operating losses, financial distress and exit, and hence, the ISP will not have sufficient incentive to remain in any coalition – the ISP will most probably shift its focus and resources to reducing its non-transmission costs, and will probably attempt to create other coalitions with third parties. In general, participation in these coalitions is not mandatory and ISPs can negotiate with various coalitions for membership or exit; subject to terms of Peering Agreements. Hence, there are three possible states which are: a) X > Y, b) Y > X, and X = Y. However, if X > Y, and $R_a, R_m \in (X,Y)$, the average coalition member will have some incentives to remain in the coalition, but such member can improve its position by negotiating with other non-member ISPs for Peering Agreements, or by expanding its own physical infrastructure or by joining third-party coalitions. Even where membership in a coalition is mandatory for all or part of the time period, coalition member ISPs can still improve their payoffs by expanding their own physical networks.

Similarly, for any time interval t and for any distance *d* in the coalition's network, if the average peer-unit Interconnection Fee ($P_a$) or Marginal per-unit Inter-connection Fee ($P_m$) (for transporting a unit of bandwidth or packet) is below a certain dollar threshold (Y), the ISP may not care much about Interconnection Fees, and will not have adequate incentives to remain in any coalition of ISPs. Also, if $P_a$ and $P_m$ are above a certain dollar threshold (X), the ISP will have strong incentives to leave the coalition and expand its own physical network; or strong incentives to reduce its physical network and enter into more Peering/Interconnection Agreements with non-coalition ISPs. In general, participation in these coalitions is not mandatory and ISPs can negotiate with various coalitions for membership or exit; subject to terms of Peering Agreements. Hence, there are three possible states which are: a) X > Y, b) Y > X, and X = Y. However, if X > Y, and $P_a$, $P_m$ $\varepsilon$ (X,Y), the average coalition member will have some incentives to remain in the coalition, but such member can improve its position by negotiating with other non-member ISPs or by expanding its own physical infrastructure or by reducing its physical network infrastructure or by joining third-party coalitions. Even where membership in a coalition is mandatory for all or part of the time period, coalition-member ISPs can still improve their payoffs by expanding their own physical networks. Thus, there can never be any Nash Equilibrium or Stackelberg Equilibrium for such coalitions of transporting ISPs.

Similarly, for any time interval *t* and for any distance *d* in the coalition's network, if the average peer-unit Revenue ($F_a$) or Marginal per-unit Revenue ($F_m$) (for transporting a unit of bandwidth or packet) is below a certain dollar threshold ($Y_f$), the ISP will not have adequate incentives to remain in any coalition of ISPs, primarily because the ISP will face financial distress and possible exit from the market. Also, if $F_a$ and $F_m$ are above a certain dollar threshold ($X_f$), the ISP will have: a) strong incentives to leave the coalition and expand its own physical network; or b) substantial incentives to remain in the coalition but vary the size of its network, and hence, the amount of its payoff, or c) substantial incentives to reduce the size of its physical network and enter into more Peering/Interconnection Agreements with both coalition-member ISPs and or non-coalition ISPs. In general, participation in these coalitions is not mandatory and ISPs can negotiate with various coalitions for membership or exit; subject to terms of Interconnection/Peering Agreements. Hence, there are three possible states which are: a) $X_f$ > $Y_f$, b) $Y_f$ > $X_f$, and $X_f$ = $Y_f$. However, if $X_f$ > $Y_f$, and $F_a$, $F_m$ $\varepsilon$ ($X_f$, $Y_f$), the average coalition member will have some incentives to remain in the coalition, but such member can improve its position by negotiating with other non-member ISPs or by expanding its own physical infrastructure or by reducing its physical network infrastructure or by joining third-party coalitions. Even where membership in a coalition is mandatory for all or part of the time period, coalition-member ISPs can still improve their payoffs by expanding their own physical networks. Thus, there can never be any Nash Equilibrium or Stackelberg Equilibrium for such coalitions of transporting ISPs. ∎

**Theorem-10**: *The Shapley Value Allocation Is Inefficient And UnFeasible Because The Shapley Value Function erroenously Assumes That The Resources Of All Coalition members Are Fixed for Any Time Interval t; and Each Coalition member's "Contribution" To The Mechanism Is Limited/finite.*

*Proof*: Each coalition member has different resources that vary in terms of availability and size; and coalition members can vary their contribution to the Shapley Value mechanism (eg. increase the physical size of their network; or change their routing algorithms). Furthermore, each coalition member's contribution to the mechanism can be infinite and unlimited – for example, each coalition member may be able to build a network that covers the subjection city or region, and maintain the network on a 24/7 basis indefinitely. ∎

**Theorem-10**: *For Any Coalition S, There Certain Conditions That Must Exist Before There Can Be A Feasible Aplication Of The Shapley Value To Allocations In Networks; But All Such Conditions Are Individually And Collectively Un-Feasible.*

*Proof:* These conditions are related to the nature of networks, and are all un-feasible and are as follows. The proof of the un-feasibility of each condition is straightforward.

***Condition #i***: *In The Shapley Value Formula, N (total number of ISPs in the market/region) and s (number of ISPs in the coalition) Are Constant For All Time Periods t And For Any Region r.* This condition is not feasible because ISPs will enter and exit the coalition over time.

***Condition #ii***: *The Marginal Contribution Of Coalition Member (ISP1) Is Equally Likely To Occur As The Marginal Contribution Of Any Other Coalition Member During Any Time Interval t.* This condition is not feasible because different ISPs have very different resources, network architectures and commitment to the coalition.

***Condition #iii***: *The terms Of The Coalition Agreement (Resource Allocation Formula) Remains Constant For Any Time Interval t.*

***Condition #iv***: *In Any Coalition S, And For Any Time Interval t, Utilities Are Always Transferable Among Any Two Pairs Of ISPs.* This condition is not feasible because different ISPs have very different resources, network architectures and commitment to the coalition. In thus context, utility is in the form of transmission and brand equity. Furthermore, the physical layout

***Condition #v***: *For any time interval t, and for any number of ISPs in a market denoted by n, and for any distance d in the coalition members' network, in the Shapley Value Formula for calculating payoffs, as s $\to$ 0 (as the number of coalition members approaches zero), the absolute effect/contribution of each coalition member-ISP remains at least constant but increases and is proportional to the ISP's network size.*

***Condition #vi***: *For any time interval t, and for any distance d in the coalition members' network, in The Shapley Value Formula for calculating payoffs, as s $\to$ n (as the number of coalition members approaches the total number of ISPs in the market), the effect/contribution of each coalition member-ISP remains at least constant and is proportional to the ISP's network size.*

***Condition #vii***: *The governance structure of the Coalition does not matter and does not affect the contributions and payoffs of coalition members.* This condition is not feasible. The Peering/Interconnection Agreement may determine the nature of the governance of the coalition, or sub-coalitions within the coalition. The coalition can be in the form of a) separate agreements among ISPs at various times, b) un-coordinated and separate strategic alliances among ISPs, c) coordinated strategic alliances among ISPs that anticipate entry/exit of alliance members, d) un-coordinated joint ventures among ISPs, c) coordinated joint ventures among ISPs that anticipate entry/exit of alliance members, f) government-mandated Interconnection Agreements. The governance includes: a) determination of costs and revenues, b) collection and sharing of revenues, c) dispute resolution methods.

***Condition #viii***: *All Interconnection Agreements (Peering Agreements And Transiting Agreements) Are The Same*.

***Condition #ix***: *Illegal Online Filesharing does not affect ISP's Profitability And Cost Structure*. This condition is not feasible because illegal online filesharing affects the profitability and QoS of ISPs. Most of the existing literature on network pricing analysis explicitly omits any analysis of the effects of illegal online filesharing.

***Condition #x***: *QoS concerns does not affect the probability of formation of Coalitions*. This condition is not feasible in most instances, because more ISPs are becoming more concerned about QoS, and are using QoS as a basis to segment their services (to offer different tiers of internet access).

***Condition #xi***: *Antitrust Laws don't matter*. On the contrary, antitrust laws preclude the types of coalitions that are envisioned or implied by Shapley Value analysis and Stackelberg Games.

**Condition #xii**. *All Coalition members "contribute" a minimum amount of resources (ie. have a physical network)*. Shapley Value analysis may be accurate only if all coalition members contribute some minimum level of resources to the mechanism/game. The Shapley Value formula for calculating the payoffs fails if one or more coalition members does not contribute any resources (eg physical network; bandwidth) to the game. On the contrary, some ISPs can be only marketing companies that don't have any physical networks but use Peering/Interconnection Agreements to transport data and thus, earn revenues.

***Condition #xiii***. *The Probabilities Of Formation of each Coalition are equal*. Shapley Value analysis erroneously assumes that the probability of each possible coalition (and probability of each member's entry into the coalition) is the same.

**Condition #xiv**: *Each Coalition Member's effort is equal to its "contribution" which in turn is equal to an equivalent amount of utility*. The Shapley Value Function Is Not Applicable To Coalition Analysis Because Each Coalition Member's effort is Not equal to its "contribution" which in turn is not equal to an equivalent amount of utility.

**Condition #xv**: *The Utility created by any Interconnection Agreement In the Coalition Is Constant for any packet-size and over any time period t, and such utility Benefits Only the two contracting ISPs*.

**Condition #xvi**: *Each ISPs contribution to the coalition Is Measured primarily by its geographic coverage and its connections to other ISPs.* ∎

**Theorem #11**. *The Shapley Value Allocation Is Not Feasible For Revenue/Cost Allocation In Networks Because Shapley Value function (for profit sharing) Does not account for differences in sources of demand among coalition members (ie. share of customers).*

*Proof*: The Shapley Value does not account for differences in sources of demand for the coalition's services and products and errornously assumes that demand is uniform and equal for all coalition members – the Sharing mechanism does not account for the number of customers delivered to the coalition by any ISP. To be accurate, the sharing mechanism should reward members for the volume of business and the quality of business (eg. long term vs. short term customers; high-bandwidth vs. low-bandwidth transmissions) that they bring to the coalition. ∎

**Theorem #12**: The *Network Transmission (And Revenue Sharing) Game Is Not A Cooperative Game*.

*Proof*: Network access transmission and revenue allocation is not a cooperative game or semi-cooperative game. The fact that ISPs enter into Interconnection Agreements doesn't make any revenue allocation system a "cooperative" game. Network access (and any implicit Revenue allocation or cost sharing) is a Non-Cooperative Game because:

**a**) In any given market, most ISPs are fierce competitors, and each ISP spends substantial amounts of money on advertising, promotions and marketing.

**b**) Each ISP spends substantial amounts of capital (cash; human capital; intellectual capital) on building its own physical network, with the intent of providing as much geographical coverage as possible. If ISPs built their networks with the objective of minimizing the size of their physical networks (and obtaining as much coverage as possible through Interconnection Agreements), then it could be inferred that the ISPs seek and heavily depend on "cooperation" although the result would likely be sub-optimal and insufficient geographical coverage of each ISP's network or of the combined networks. Thus, it can be reasonably inferred that ISPs enter into Interconnection Agreements as a matter of convenience, and to accommodate their customers needs (if and when the ISP identifies certain traffic patterns). Furthermore, of the two types of Interconnection Agreements, Peering Agreements) account for more than Seventy Percent of all Interconnection Agreements, and Peering Agreements do not result in any additional costs for the ISP (each ISP agrees to carry the other ISP's traffic for free).

**c)** In most markets, most consumers have only one or two ISPs, and one internet access point. Hence, the consumer's choice of one ISP results in an automatic economic loss for other ISPs in the same market. This economic loss occurs because of the sunk costs (physical infrastructure; advertising costs; administrative costs) incurred by ISPs.

**d)** Each time that an ISP (ISP1) transits a packet for another coalition member (ISP2), ISP1 losses brand equity and ISP2 gains brand equity because ISP2 is completing service. Interconnection Fees often don't compensate ISPs for such loss of brand equity.

Therefore, Shapley Value Function is not applicable to access pricing. ∎

Properties Of An Allocation Mechanism

Furthermore, the "Properties" of the allocation mechanism that were stated in Ma et. al. (2009) are not valid because of the following reasons:

a) *Symmetry* – this definition of symmetry in Ma, et al (2009) is not an accurate definition of symmetry within and outside the context of Networks, and it erroneously assumes that any two ASes (ISPs) that have the same "contribution" are in the same circumstances, and or have the same cost structure and generate the same utility. Here, the distinctions among "contribution" and "utility" and "marginal utility" are critical, because "contribution" does not necessarily result in a change in utility or marginal utility.

b) *Dummy*: As defined, this property ignores the economic value of flexibility. A Dummy AS that provides utility in some but not all transactions should recive some share of the profits. Even completely–Dummy have value because they provide flexibility in case of emergencies.

c) *Strong Monotonicity*: This property is not valid in all circumstances because: a) "contribution" does not necessarily equate to added utility, b) there may be decreasing returns to scale wherein the utility created by the ISP's efforts declines as efforts increase.

d) *Additivity*: This is not a valid property in all circumstances because there may be increasing returns-to-contribution, or decreasing-returns-to-contribution for some members or each member or all members of the Coalition.

Consider a sub-game in which there is a market and there are six ISPs in the market $ISP_1$........$ISP_6$. D% of total traffic has to pass though some part of each ISP's network in order to complete the transmission. Each ISP has its own FLCC ($A_n$) and so there are $A_1$............$A_6$ FLCCs. The degree of similarity and or goal congruence between any two pairs of FLCC algorithms is referred to as Congruency Index (CI) which increases as the two FLCC's objectives and criteria are more similar. For traffic to flow adequately, there must be at least B% of CI among FLCC's $A_1$.............$A_6$. If CI is below B%, then there is substantial risk of packet delay or packet loss as B tends to 100%, the probability of successful transmission tends to one. FLCCs are typically highly proprietary algorithms. While technology companies typically meet to set industry standards, such standards are typically for non-proprietary items or processes. This implies that there must be some minimum amount of collusion or collaboration among ISPs in order to ensure that there is the minimum CI percentage (B%) – such collusion is most probably illegal under most antitrust laws.

## Conclusion

The Shapley Value and related theories are clearly not applicable in internet Network pricing analysis. The elimination or relaxation of some of the prerequisite constraints implicit in Shapley Value analysis will render the results meaningless. Antitrust laws also preclude applicability of Shapley-Value solutions. Thus, most of the theories introduced in Cao, Shen, Milito & Wirth (2002); Ma, Chiu, Lui, Misra & Rubenstyein (2009); and Perez-Castrillo & Wettstein (2001), are wrong. For the same and similar reasons stated in this article, and contrary to Lima, Contreras & Padilha-Feltrin (2008); Jia & Yokoyama (2003); and Kattuman, Green & Bialek (2004), the Shapley Value Function cannot be used for allocating electricity transmission costs; and cannot be used for allocating costs in Networks (Roth & Verrecchia (1979)).

J. Chuang, Economics of Ad-Hoc, Overlay and Traditional Networks: A Topological Perspective, Position Paper for the ACM SIGCOMM Workshop on Network Research: Exploration of Dimensions and Scope, Karlsruhe Germany, August 2003.

P. Laskowski, J. Chuang. Antitrust Scrutiny of Price-Fixing Clauses in Patent Licenses. Working Paper, 2005.

J. Grossklags, N. Christin, J. Chuang. Security Investment (Failures) in Five Economic Environments: A Comparison of Homogeneous and Heterogeneous User Agents. 7th Workshop on the Economics of Information Security (WEIS'08), June 2008.28